\documentclass[12pt]{iopart}
\usepackage{graphics}
\def\prl{{\em Phys. Rev. Lett. }}
\def\prc{{\em Phys. Rev. {\bf C} }}

\def\epja{{\em Eur. Phys. J. {\bf A}}}
\def\plb{{\em Phys. Lett. {\bf B}}}

\def\zpc{{\em Z. Phys. {\bf C}}}

\def\app{{\em Acta Physica Polonica {\bf B}}}

\begin{document}
\title{Saturation of $E_T/N_{ch}$ and Freeze-Out Criteria in Heavy-Ion Collisions}
\author{J Cleymans$^1$, R Sahoo$^{2,3}$\footnote{Email: Raghunath.Sahoo@cern.ch, Speaker} , D P Mahapatra$^2$, D K Srivastava$^4$ 
and S Wheaton$^1$}
\address{$^1$UCT-CERN Research Centre and Department  of  Physics,University of Cape Town, 
Rondebosch 7701, South Africa}
\address{$^2$Institute of Physics, Sachivalaya Marg, Bhubaneswar 751005, India}
\address{$^3$SUBATECH, 4, Rue Alfred Kastler, BP 20722 - 44307 Nantes Cedex 3, France}
\address{$^4$Variable Energy Cyclotron Centre, 1/AF Bidhan Nagar, Kolkata 700064, India}

%
\begin{abstract}
The pseudorapidity densities of transverse energy, the charged particle
multiplicity and their ratios, $E_T/N_{\textrm{ch}}$, are estimated at mid-rapidity, 
in a statistical-thermal model based on chemical freeze-out criteria, for a 
wide range of energies from GSI-AGS-SPS to RHIC. It has been observed that 
in nucleus-nucleus collisions, $E_T/N_{\textrm{ch}}$ increases rapidly with beam energy 
and remains approximately constant at about a value of 800 MeV for beam energies 
from SPS to RHIC. $E_T/N_{\textrm{ch}}$ has been observed to be almost independent of 
centrality at all measured energies. The statistical-thermal model describes the 
energy dependence as well as the centrality independence, qualitatively well. 
The values of $E_T/N_{\textrm{ch}}$ are related to the chemical freeze-out criterium, 
$E/N \approx 1~GeV$  valid for primordial hadrons. We have 
studied the variation of the average mass $(<$\textrm{MASS}$>), 
N_{\textrm{decays}}/N_{\textrm{primordial}}, N_{\textrm{ch}}/N_{\textrm{decays}}$ 
and $E_T/N_{\textrm{ch}}$ with $\sqrt{s_{NN}}$ for all freeze-out criteria discussed 
in literature. These observables show saturation around SPS and higher $\sqrt{s_{NN}}$, 
like the chemical freeze-out temperature ($T_{\textrm{ch}}$). 
\end{abstract}
%
\maketitle 
%
\section{Introduction}
The final state particles in relativistic heavy-ion collisions hardly remember
about their primordial origins, as they are subjected to many rescatterings in the
hadronic stage. This has given rise to the interpretation of hadron production
in terms of thermal and statistical models which assume chemical and kinetic
freeze-out of the particles. All relativistic heavy-ion experiments have so 
far confirmed the validity of $E/N \approx$ 1 GeV as a freeze-out criterium,  
with $E$ and $N$ being, respectively the total energy and  particle number 
of the primordial hadronic resonances before they decay into stable hadrons.
These quantities cannot be determined directly from experiment unless the 
final state multiplicity 
is  low and hadronic resonances can be identified, which is not the case in  
relativistic heavy-ion collisions. It is thus not straightforward to link 
$E/N$ to directly measurable quantities. In this paper, we establish an 
approximate connection between $E/N$ and the ratio of the pseudo-rapidity 
density of transverse energy and that of the charged particle yield, 
[$(dE_T/d\eta)/(dN_{\textrm{ch}}/d\eta) \equiv E_T/N_{\textrm{ch}}$], at mid-rapidity,
for  beam energies ranging from about 1 AGeV up to 200 AGeV. In this energy range,
$E_T/N_{\textrm{ch}}$ at first increases rapidly from SIS to AGS, then saturates 
to a value of about 800 MeV at SPS energies and remains constant
up to the highest available RHIC energies~\cite{raghu}. The present analysis 
of $E_T/N_{\textrm{ch}}$ uses the hadron resonance gas 
model (thermal model). Our analysis starts by relating the number of charged 
particles seen in the detector to the number of primordial hadronic resonances 
and the transverse energy to the energy $E$ of primordial hadrons. The present 
status of $E/N$ could be found in Ref. ~\cite{raghu1}. 

\begin{figure}
\begin{center}
\resizebox{0.95\columnwidth}{!}{%
\includegraphics{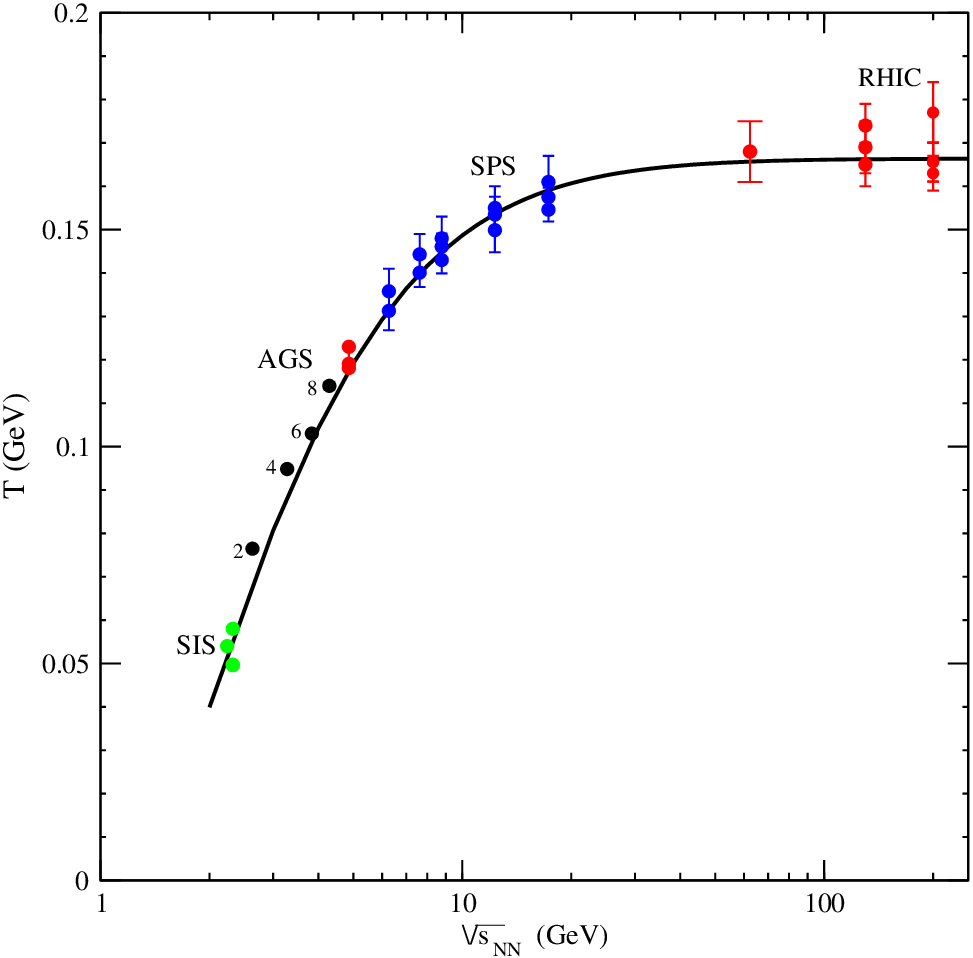}
\includegraphics{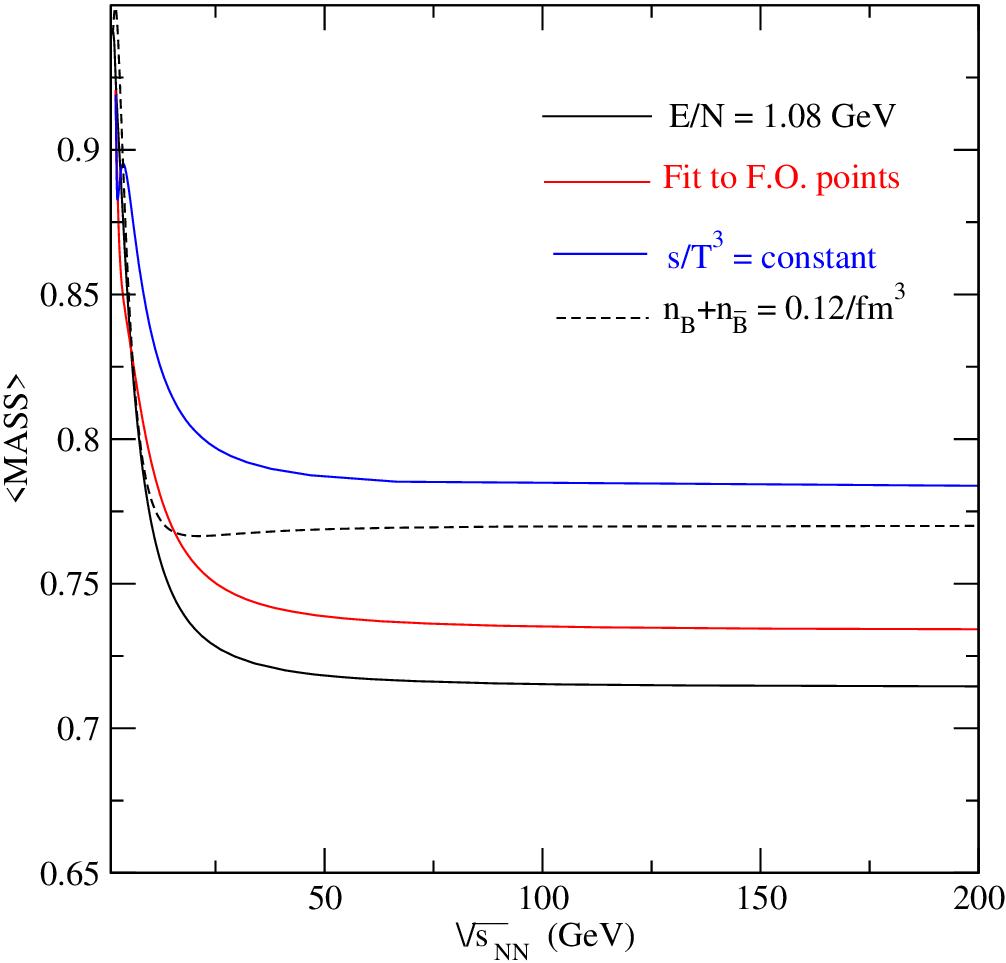}}
\caption{(a) Saturation of the chemical freeze-out temperature at high energies, 
(b) Saturation of the average mass in the hadronic resonance 
gas model at high beam energies for various freeze-out criteria 
proposed in the literature~\cite{PBMS,Majiec,TawfikEntropy,tawfik2}.
}

\label{meanmass}
\end{center}
\end{figure}

   In this paper, all thermal model calculations were performed using the 
THERMUS package~\cite{THERMUS}. 
%
%
At high energies the chemical freeze-out temperature saturates at
a value of about 160 - 170 MeV as shown in Figure~\ref{meanmass}(a) 
and at the same time the baryon chemical potential becomes very 
small~\cite{FreezeOuts}. As a consequence, several other quantities 
also become independent of beam energy. The average mass of hadronic 
resonances saturates at approximately the $\rho$ mass at high energies 
as shown in Figure~\ref{meanmass}(b). The ratio of all hadrons after resonance 
decays to the number of directly emitted hadrons at chemical freeze-out
saturates at a value of about 1.7 as shown in Figure~\ref{joint1}. All of 
these are direct consequences of the saturation of the freeze-out 
temperature observed in Figure~\ref{meanmass}(a) for increasing beam 
energies and the associated convergence of the baryon chemical potential 
to zero. 

\section{Freeze-out}
A theoretical description of the whole duration of evolution of the fireball 
produced in heavy ion collisions is difficult as different degrees of 
freedom are important at various stages of the evolution. The thermal model uses
the hadronic degrees of freedom at the latest stages of the evolution of the
fireball when the chemical composition of different particle species stops 
changing (chemical freeze-out) and then the particle mean free path becomes 
larger than the system size and thus the system is frozen kinetically 
(thermal freeze-out). Freeze-out could be a complicated process involving 
duration in time and a hierarchy where different kinds of particles and 
reactions switch-off at different times, giving rise to the concept of
{\it ''differential freeze-out''}. By kinetic arguments, it is expected 
that the reactions with lower cross-sections switch-off at higher 
densities/temperatures compared to reactions with larger cross-sections. 
Hence, the chemical freeze-out (corresponds to inelastic reactions) 
occurs earlier in time compared to the kinetic freeze-out (corresponds to 
elastic reactions). In line to this argument, one can think of strange or charmed
particles decoupling from the system earlier than the lighter hadrons.
A series of freeze-outs could be imagined corresponding to particular reaction 
channels ~\cite{baran}. However, we will focus on chemical and kinetic freeze-outs 
in our discussions.

\section{Results and Discussions}
The transverse energy density in pseudorapidity, $dE_T/d\eta \equiv E_T$, 
has two components, the hadronic one, $E_T^{\textrm{had}}$, and the 
electromagnetic one, $E_T^{\textrm{em}}$, coming from the electromagnetic 
particles (photons, electrons and positrons). Electromagnetic calorimeters 
are used to measure $E_T^{\textrm{em}}$, whereas hadronic calorimeters or 
the Time Projection Chamber with experimental correction for long-lived 
neutral hadrons are used to measure $E_T^{\textrm{had}}$~\cite{star200GeV,raghu2}. 
The energy of a particle is defined as being the kinetic energy for nucleons, 
for anti-nucleons as the total energy plus the rest mass and for all other 
particles as the total energy~\cite{star200GeV,helios}. 

Experiments  
have reported a constant value of the ratio $E_T/N_{\textrm{ch}}\sim 0.8$ GeV from 
SPS to RHIC~\cite{star200GeV,phenixSyst}, with the ratio being almost independent 
of centrality of the collision for all measurements at different energies. 
In all cases, the value of $E_T/N_{\textrm{ch}}$ has been taken for the 
most central collisions at mid-rapidity. At the end of this paper we consider 
the centrality dependence of  $E_T/N_{\textrm{ch}}$. When this ratio is observed 
for the full range of center of mass energies, it shows two regions~\cite{phenixSyst}.
In the first region from lowest $\sqrt{s_{NN}}$ to SPS energy, there is a steep 
increase of the $E_T/N_{\textrm{ch}}$ ratio with $\sqrt{s_{NN}}$. In this regime, 
the increase of $\sqrt{s_{NN}}$ causes an increase in the $\left<m_T\right>$
of the produced particles. In the second region,  SPS  to higher energies, the 
$E_T/N_{\textrm{ch}}$ ratio is very weakly dependent on $\sqrt{s_{NN}}$. 

\begin{figure}
\begin{center}
\resizebox{0.75\columnwidth}{!}{%
\includegraphics{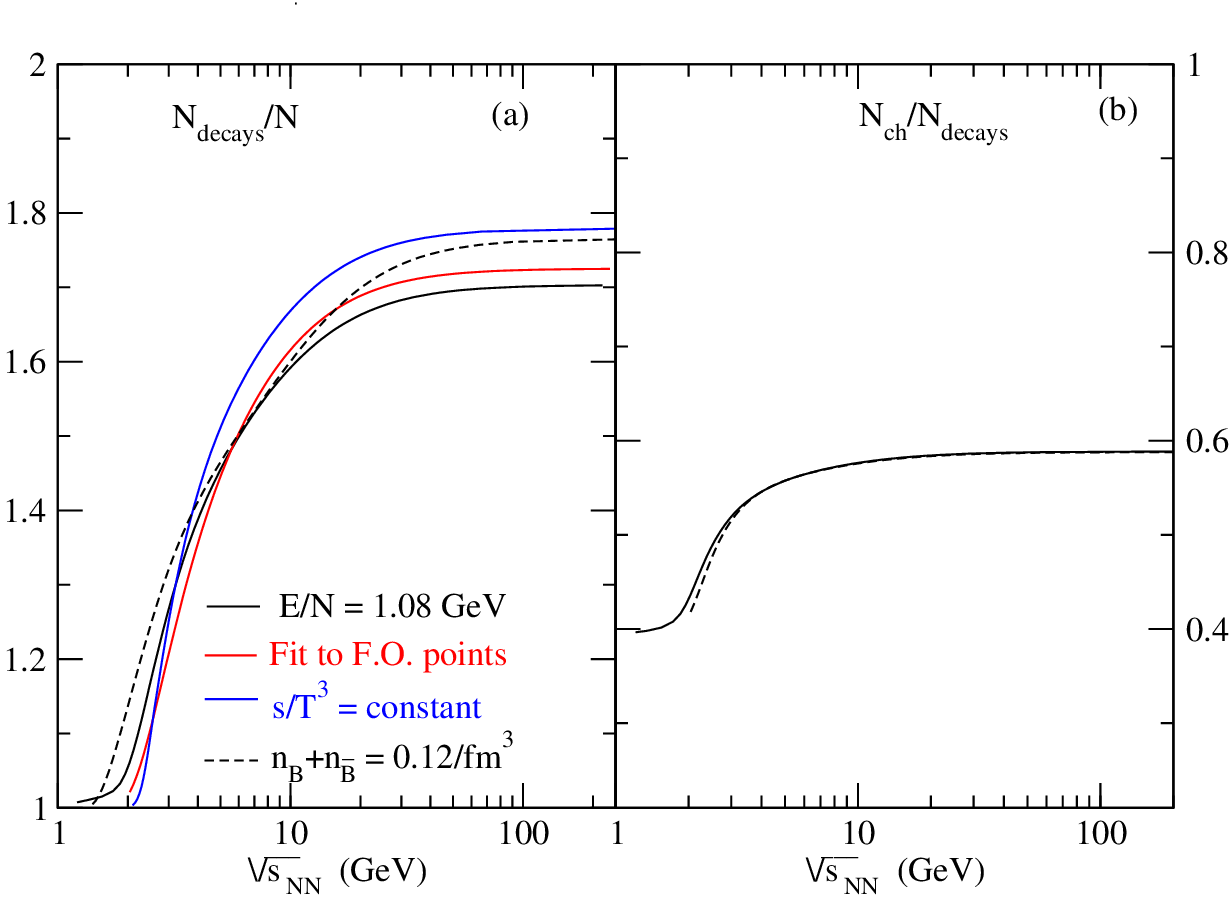}}
\caption{Saturation of $N_{\textrm{decays}}/N$ (a) and 
$N_{\textrm{ch}}/N_{\textrm{decays}}$ (b) with $\sqrt{s_{NN}}$. 
In (a) the results from various freeze-out criteria are indicated.
In (b) the different freeze-out criteria give results that are indistinguishable.}
\label{joint1}
\end{center}
\end{figure}

To estimate $E_T/N_{\textrm{ch}}$ in the thermal model, we relate the number of 
charged particles, $N_{\textrm{ch}}$, to the number, $N$, of primordial hadrons.
To estimate the charged particle multiplicity at different center of mass energies 
from the thermal model, we proceed as follows. First we study the variation of the
ratio of the total particle multiplicity in the final state, $N_{\textrm{decays}}$, 
and that in the primordial i.e. $N_{\textrm{decays}}/N$ with $\sqrt{s_{NN}}$.
This ratio starts from one, since there are only a few resonances produced at 
low beam energy and becomes almost independent of energy after SPS energy with
a saturated value of around 1.7. The excitation function of $N_{\textrm{decays}}/N$
is shown in Figure~\ref{joint1}(a). Secondly, we have studied the variation of the 
ratio of charge particle multiplicity and the particle multiplicity in the final 
state ($N_{\textrm{ch}}/N_{\textrm{decays}}$) with $\sqrt{s_{NN}}$. 
This is shown in Figure~\ref{joint1}(b). The $N_{\textrm{ch}}/N_{\textrm{decays}}$ 
ratio starts around 0.4 at lower $\sqrt{s_{NN}}$ and shows an energy independence 
at SPS and higher energies. At lower SIS energy, the baryon dominance at 
mid-rapidity makes $N_{\textrm{ch}}/N_{\textrm{decays}}\sim N_{\textrm{proton}}
/N_{\textrm{(proton+neutron)}}$ which has a value of 0.45 for  Au-Au collisions.

As the next step, we connect the transverse energy $E_T$ to the the energy of 
the primordial hadrons $E$. In the hadronic resonance gas model there is a sum 
over all hadrons; furthermore, taking into account the experimental configuration 
which leads to adding the mass of the nucleon for anti-nucleons and subtracting 
the same for nucleons one has
\begin{eqnarray}
\left<E_T\right> & \equiv & V\sum_{i={\textrm{nucleons}}}\int\frac{d^3p_i}{(2\pi)^3} (E_i-m_N)\sin\theta_i~~f(E_i) \nonumber\\
&&+V\sum_{i=\textrm{anti-nucleons}}\int 
\frac{d^3p_i}{(2\pi)^3} (E_i+m_N) \sin \theta_i~~f(E_i)\nonumber\\
&&+V\sum_{i=\textrm{all~others}}\int \frac{d^3p_i}{(2\pi)^3} E_i \sin \theta_i~~f(E_i) , \nonumber \\
&= &  \frac{\pi}{4}\left[\left<E\right>-m_N\left<N_B-N_{\bar B}\right>\right] .
\end{eqnarray}

Here $\theta$ is the polar angle of a particle with the beam direction and 
$f(E)$ is the statistical distribution factor. The above equation relates the 
transverse energy measured from the data and that estimated from the thermal model. 
In the limit of large beam energies one has
\begin{eqnarray}
\lim_{\sqrt{s_{NN}}\rightarrow\infty}{\left<E_T\right>\over N_{\textrm{ch}}} 
&=&{\left<E_T\right>\over 0.6 N_{\textrm{decays}}},\nonumber\\
&=&{\pi\over 4}{1\over 0.6}{E\over 1.7 N} ,\nonumber\\
&=&0.77{E\over N},\nonumber\\
&\approx& 0.83~\textrm{GeV} . 
\end{eqnarray}
This value of $E_T/N_{\textrm{ch}}$ is close to the values measured at RHIC energies and is 
independent of collision species, centrality of the collisions  and $\sqrt{s_{NN}}$. 
It should be noted that the measured $E_T$ will be affected by the transverse 
collective flow and by the difference between chemical freeze-out and kinetic 
freeze-out temperatures and therefore the  description presented here is only a 
qualitative one. 
An analysis including flow was presented in Figure 17 of the review article by 
Kolb and Heinz~\cite{heinz} who show that this improves the agreement with 
the data at SPS and RHIC beam energies.
A detailed comparison  in the framework of a specific model with a single 
freeze-out temperature, has been made in Ref.~\cite{prorok}.

At higher energies, when $\mu_B$ nearly goes to zero, the transverse energy 
production is mainly due to the meson content in the matter. The intersection 
points of lines of constant $E_T/N_{\textrm{ch}}$ and the freeze-out line give 
the values of $E_T/N_{\textrm{ch}}$ at the chemical freeze-out. Hence at 
freeze-out, given the values of $E_T/N_{\textrm{ch}}$ from the experimental 
measurements we can determine $T$ and $\mu_B$ of the system~\cite{raghu}.

\begin{figure}
\begin{center}
\resizebox{0.95\columnwidth}{!}{%
\includegraphics{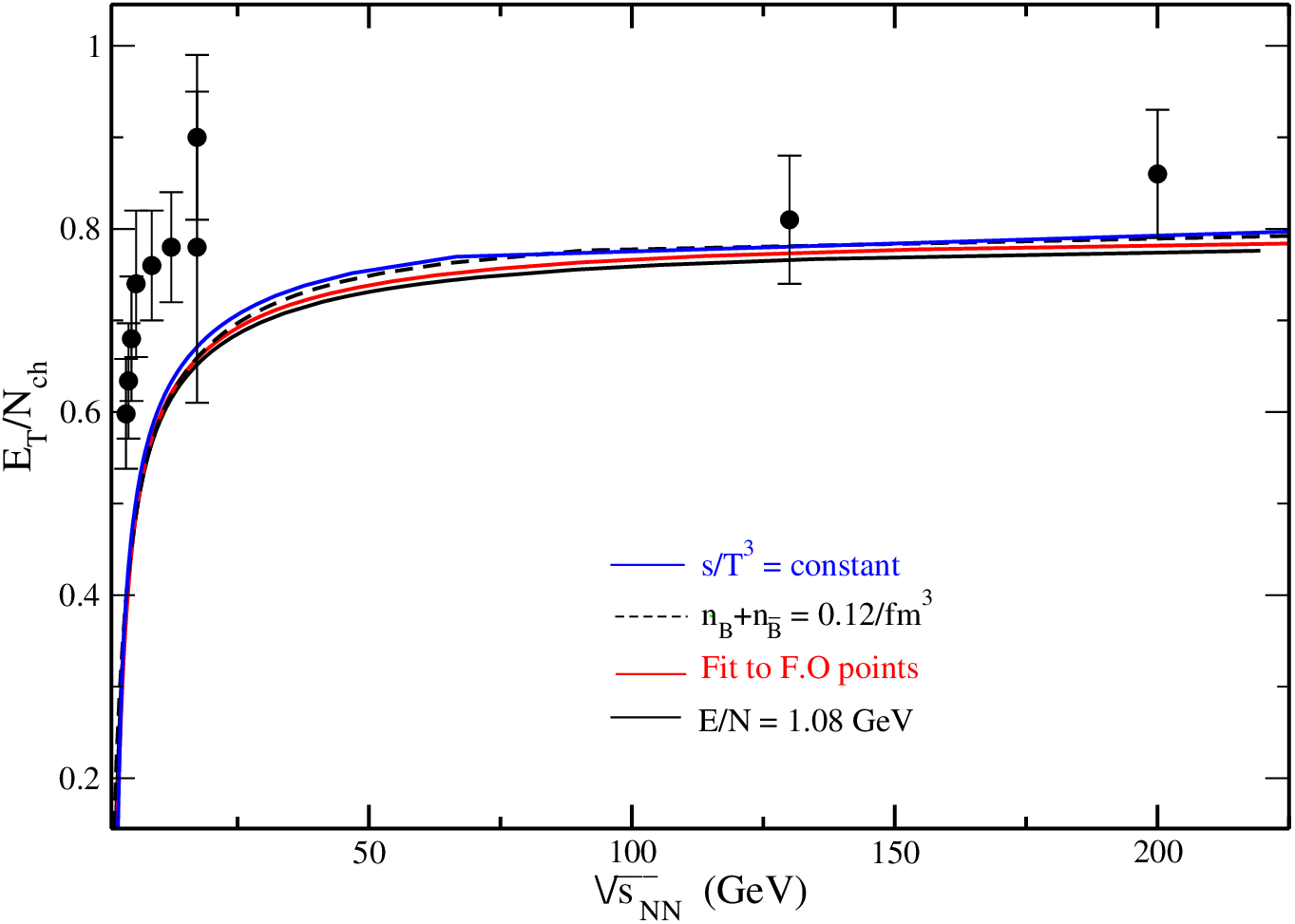}
\includegraphics{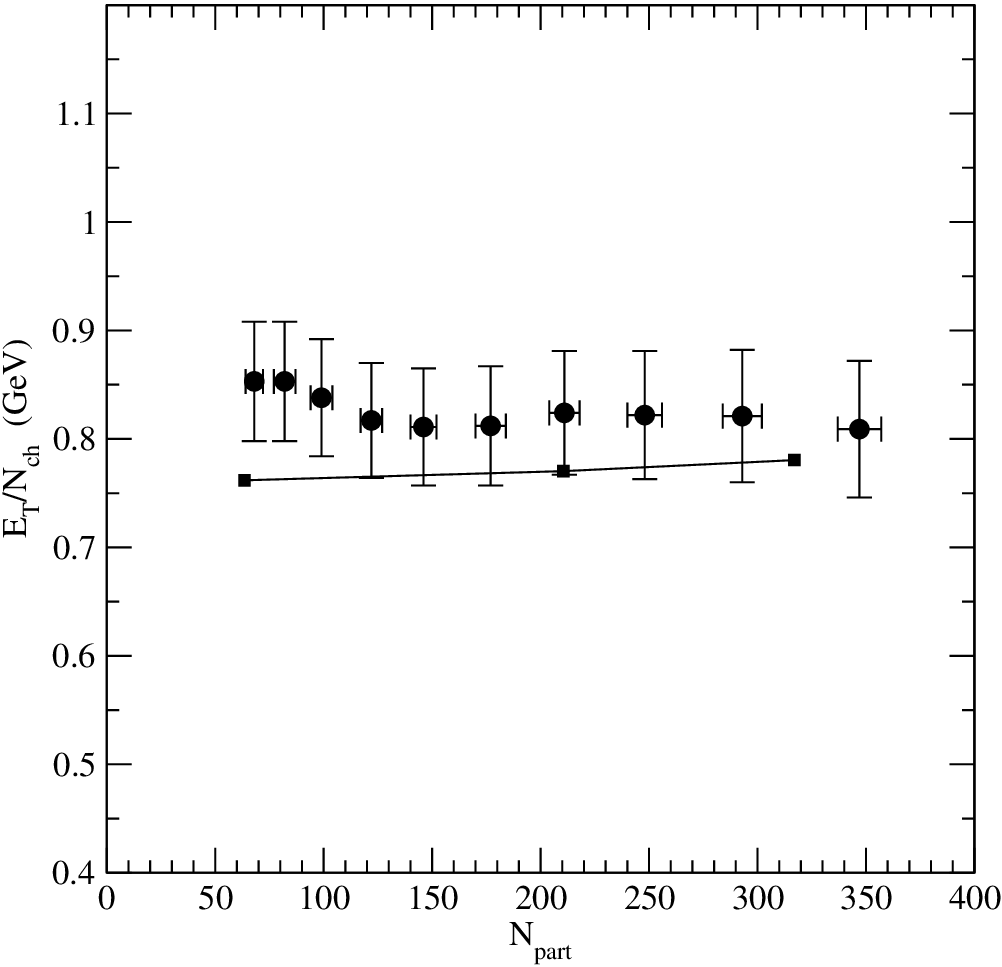}}
\caption{(a) Comparison between experimental data for
 $E_T/N_{\textrm{ch}}$ with $\sqrt{s_{NN}}$ and the thermal model
using  $E/N = 1.08$ GeV as  well as other  freeze-out 
conditions~\cite{PBMS,Majiec,TawfikEntropy,tawfik2}, (b) The variation of 
$E_T/N_{\textrm{ch}}$ with $N_{part}$ for 130 GeV Au+Au collisions at RHIC 
~\cite{phenix130GeV} with corresponding thermal model estimates.}
\label{etNchCM}
\end{center}
\end{figure}

For the most central collisions, the variation of $E_T/N_{\textrm{ch}}$ 
with center of mass energy is shown in Figure~\ref{etNchCM}(a). The data have been 
taken from Ref.~\cite{raghu},
and are compared with the corresponding calculation
from the thermal model with chemical freeze-out. We have checked explicitly
that other freeze-out criteria discussed in the literature give almost identical 
results for the behavior of $E_T/N_{\textrm{ch}}$ as a function of $\sqrt{s_{NN}}$;
this is the case for the fixed baryon plus anti-baryon density condition~\cite{PBMS}
and also for fixed normalized entropy density condition,
$s/T^3$ = 7~\cite{Majiec,TawfikEntropy,tawfik2}.
As is shown in Figure~\ref{etNchCM}(b), centrality behavior of $E_T/N_{\textrm{ch}}$ is well 
reproduced by the thermal hadronic resonance gas model.

Taking the arguments by Heinz {\it et al.} ~\cite{heinz1}, if the freeze-out 
is a kinetic process, it is controlled by the competition between local 
scattering (moving the system towards equilibrium) and global expansion 
(driving the system out of equilibrium). The resulting freeze-out temperature
is therefore sensitive to the fireball expansion rate which depends on collision
centrality. Hence the kinetic decoupling temperature ($T_{\textrm{kin}}$) should depend
on centrality. Such an centrality dependence has been observed for $T_{\textrm{kin}}$,
whereas $T_{\textrm{ch}}$ has been observed to be independent of centrality ~\cite{IPSTAR}. 
The centrality independence of $T_{\textrm{ch}}$ has been interpreted as due to the chemical
decoupling of hadron abundances being driven by a phase transition during which the
chemical reaction rates decrease abruptly, leaving the system in a chemically
frozen-out state at the end of the transition. Thus we get a universal $T_{ch}$
which is insensitive to the collective dynamics but depends on the thermodynamic
parameters of the phase transition. The observation of $E_T/N_{\textrm{ch}}$ saturating
at a universal value and being independent of centrality of the collisions
could also be related to the quark-hadron phase transition through chemical
freeze-out. The saturation value of $E_T/N_{\textrm{ch}}$ can also be taken as its value 
for a pre-hadronic state, as $T_{\textrm{ch}} \sim T_C$. Irrespective of the initial 
conditions (controlled by system size and beam energy), at higher energies, the 
system evolves to the same chemical freeze-out condition.

\section{Summary}
In conclusion, we have discussed the connection between  $E_T/N_{\textrm{ch}}$  
and the ratio of primordial energy to primordial particle multiplicity, $E/N$,  
from the thermal model. This model, when  combined with  chemical freeze-out 
criteria explains the data over all available measurements for the $\sqrt{s_{NN}}$ 
and centrality behavior of $E_T/N_{\textrm{ch}}$. $E_T/N_{\textrm{ch}}$ being 
realted to $T_{\textrm{ch}}$ is associated with the quark-hadron phase transition.  
It has to be noted that  variables like $E_T/N_{\textrm{ch}}$, the chemical 
freeze-out temperature $T_{\textrm{ch}}$, 
$N_{\textrm{decays}}/N_{\textrm{primordial}}$ and 
$N_{\textrm{ch}}/N_{\textrm{decays}}$ discussed in this paper, show  saturation 
starting at SPS and continuing to higher center of mass energies. 

\ack
Three of us (JC,~RS,~DKS) would like to acknowledge the financial support of
the South Africa-India Science and Technology agreement.


\section*{References}
\begin{thebibliography}{90}

%
\bibitem{raghu} Cleymans J, Sahoo R, Mahapatra D P, Srivastava D K and Wheaton S \\ 
2008 \plb {~\bf 660} 172 and references therein
\bibitem{THERMUS} Wheaton S and Cleymans J 2005 {\it Journal of Physics G} {\bf 31} 
S1069, {\it Preprint} hep-ph/0407174
\bibitem{FreezeOuts} Cleymans J, Oeschler H, Redlich K and Wheaton S 2006 \prc {\bf 73} 
034905
\bibitem{raghu1} Cleymans J, Sahoo R, Srivastava D K and Wheaton S  
{\it Preprint} hep-ph/0712.1671, \\To appear in {\it Eur. Phys. Journal C} 
\bibitem{PBMS} Braun-Munzinger P and Stachel J 2002 {\it J. Phys. G: Nucl. Part. Phys.} 
{\bf 28} 1971
\bibitem{Majiec} Cleymans J, Stankiewicz M, Steinberg P and Wheaton S,  
{\it Preprint} nucl-th/0506027
\bibitem{TawfikEntropy} Tawfik A 2005 {\em J. Phys. G: Nucl. Part. Phys.} {\bf 31} S1105 
\bibitem{tawfik2} Tawfik A  {\it Preprint} hep-ph/0507252 and {\it Preprint} hep-ph/050824
\bibitem{baran} Baran A, Broniowski W and  Florkowski W 2004 \app {\bf 35} 779
\bibitem{star200GeV} Adams J {\it et al}. (STAR Collaboration), 2004 \prc{\bf 70} 054907
\bibitem{raghu2} Sahoo R (STAR Collaboration) 2007 {\it Transverse energy measurement 
and fluctuation studies in ultra-relativistic heavy ion collisions},  Ph.D. Thesis
\bibitem{helios} Akesson T {\it et al}. (HELIOS Collaboration), 1988 \zpc{\bf 38} 383
\bibitem{phenixSyst} Adler S S {\it et al}. (PHENIX Collaboration), 2005 \prc{\bf 71} 034908
\bibitem{phenix130GeV} Adcox K {\it et al}. (PHENIX Collaboration), 2001 \prl{\bf 87} 052301

%
\bibitem{UnifiedFO} Cleymans J and Redlich K  1998 \prl{\bf 81} 5284;
1999 \prc {\bf 60} 054908
\bibitem{heinz} Kolb P F and Heinz U  {\it Quark-Gluon Plasma}, p 634
(World Scientific, Singapore) \\Eds. R.C. Hwa and X.N. wang
\bibitem{prorok} Prorok D 2007 \prc {\bf 75} 014903, 2005 \epja {\bf 24} 93; 
\\ 2003 {\it Acta Phys. Pol. B} {\bf 34} 4219
\bibitem{heinz1} Heinz  U and Kestin G {\it Preprint} nucl-th/0709.3366
\bibitem{IPSTAR} Adams J {\it et al}. (STAR Collaboration), 2004 \prl{\bf 92} 112301

\end{thebibliography}
\end{document}